# Progress in neural networks for EEG signal recognition in 2021

Rakhmatulin Ildar, PhD

*South Ural State University, Department of Power Plants Networks and Systems, Chelyabinsk city, Russia*

ildar.o2010@yandex.ru

**Abstract**

In recent years, neural networks showed unprecedented growth that ultimately influenced dozens of different industries, including signal processing for the electroencephalography (EEG) process. Electroencephalography, although it appeared in the first half of the 20th century, was not changed the physical principles of work to this day. But signal processing technology made significant progress in this area through the use of neural networks. But many different models of neural networks complicate the process of understanding the real situation in this area. This manuscript summarizes the current state of knowledge on this topic, summarizes and describes the most significant achievements in various fields of application of neural networks for processing EEG signals. We discussed in detail the results presented in recent research papers for various fields in which EEG signals have been involved. We also examined in detail the process of extracting features from EEG signals using neural networks. In conclusion, we have provided recommendations for the correct demonstration of research results in manuscripts on the subject of neural networks and EEG.

## 1. Introduction

EEG signals for the diagnosis of diseases in neurological diagnostics read since the first half of the 20th century. Research in this area still not stopped to this day. Now the spectrum of research of EEG signals was expanded significantly. A huge number of patterns between EEG signals and motor activity, mental state, and mental activity were revealed. However, despite significant progress, it remains unknown how much information can be decoded from the EEG. Science in this direction is still in the initial path of its formation. Although the physics process of the appearance of signals in the cerebral cortex is well researched and many researchers as and Neelam [1] described in detail the movement process for neurons. Now known that when neurons are activated, synaptic currents begin to be generated in dendrites, and the current generates a magnetic field above the scalp and creates potential which recorded from the scalp in microvolts (μV). According to the frequency band, these EEG signals are divided into categories. There are 4 frequency bands of human EEG waves (figure 1, a):

- Delta: 3 Hz or lower. Delta waves are the slowest wave and have the largest amplitude. Usually, at the 3rd and 4th stages of sleep and young children of the first year, this signal band is dominant;

- Theta: with a frequency range of 3.5 Hz to 7.5 Hz. This signal is noticeable in children under 13 years of age;
- Alpha: frequency range from 7.5 to 13 Hz. Typically, the posterior regions of the head emit alpha signals. This is the main type of signal in a normal adult in a relaxed state;
- Beta: frequency greater than 14 Hz. It is markedly distinguished in patients who are alert for a long time or worry.

Another important factor in research is the commitment to the international system of electrode placement "10-20", (figure 1, b).

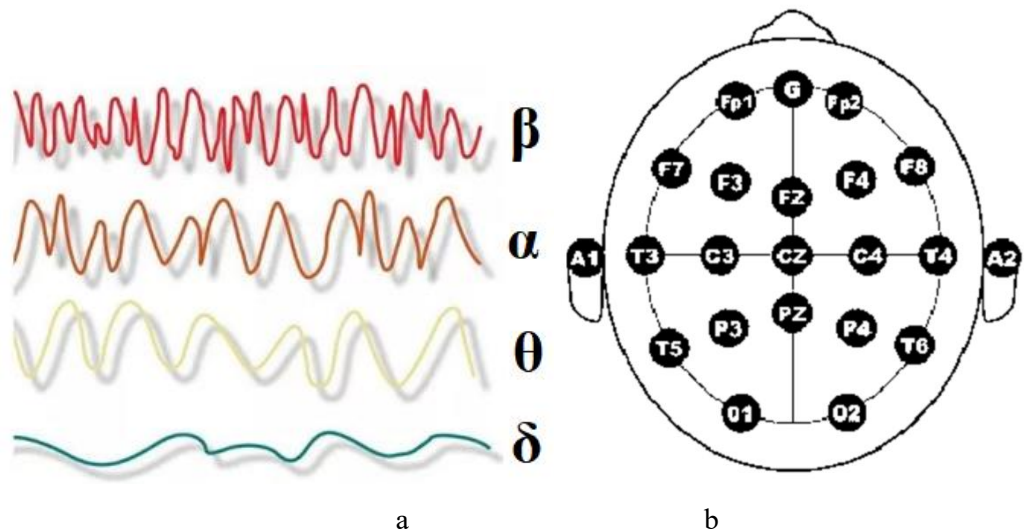

a  b

Fig.1. a - brain rhythms (β, α, θ, γ), b – the international system of placement of electrodes "10-20"

Since the data recorded by the EEG is a complex waveform, good signal processing methods are required to obtain data from the EEG. Ravan [2] described why collected data from the EEG are not used for analysis in its original form and the EEG signals are not processed in their natural state. Before proceeding to the process of extracting signs, it is better to pre-process the EEG signals. The most popular method for signal processing is the Fast Fourier Transform (FFT) method. It should be noted that there are many signal processing methods: spectral, mapping, morphological localization, period metric, correlation, auxiliary, segment analysis, and others. The use of neural networks involved in the field of EEG signal processing in this work in the following directions was distributed, figure 2.

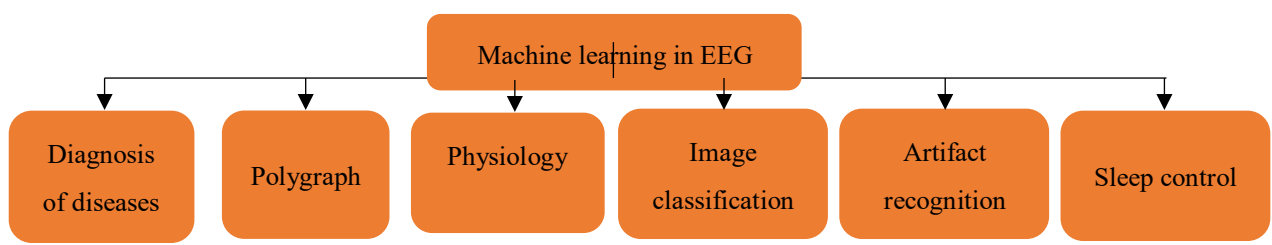

Fig.2. Classification of the directions of use of neural networks in the processing of EEG signals

We considered only completed studies in which the test results neural networks and of dry, non-contact EEG sensors and wet electrodes were used. According to their characteristics, these devices should have been suitable for clinical trials. Search for articles by keywords was carried out in the following publishers: Elsiver, Taylor & Francis, Springer, Wiley, Informa. Keyword searches were also done on the Google search engine and scholar.google.com for the last 10 years. We tried to focus on modern research. If the results coincided, an earlier source was taken for review.

The next articles are close to this manuscript. Amorim et al. [3] presented, the classification of EEG data using machine learning methods. Despite the claimed title, the theory is presented for the most part with a description of the algorithms and models of online learning that can be used in this case, references to real data are minimized. Congedo et al. [4], Sharmila et al. [5], Usman et al. [6], and Clarke et al. [7] described in the manuscripts the methods of applying neural networks without any systematic approach. This makes it difficult to imagine a complete picture of the position of machine learning in the EEG industry.

## 2. Review of EEG feature selection by neural networks

Any research in the field in which EEG signals are involved begins with EEG signal processing and feature extraction. Feature extraction is the process of extracting useful information from an electroencephalogram (EEG) signal. This is necessary to represent the correct dataset before performing the classification procedure. There are several works that are devoted to this topic. Al-Fahum et al. [8] considered the following feature extraction methods: Time Frequency Distribution (TFD), Fast Fourier Transform (FFT), Eigenvector Methods (EM), Wavelet Transform (WT), and Autoregressive Method (ARM). But the authors of the article have rather briefly described the work in which these methods are involved. Mane et al. (2019) [9], similarly to the author in the article described above, different methods of feature extraction are compared, such as wavelet transform, independent component analysis, principal component analysis, autoregressive model, and empirical mode decomposition. Raut [10] in his article considered three subsets of the obtained characteristics using the track extraction method, wavelet transform, and fractional Fourier transform. Raut compared efficiency by classification using support vector machines. In this manuscript we will briefly give a description of the main methods from the point of view of using neural networks.

### 2.1 EEG feature extraction by Fourier transforms

The Fourier transform is a family of mathematical methods based on the expansion of the original continuous function of time into a set of fundamental harmonic functions (which are sinusoidal functions) of various frequencies, amplitudes and phases. The main idea of the transformation is that any function can be represented as an infinite sum of sinusoids, each of which will be characterized by its own amplitude, frequency and initial phase. Al-Salman et al. [11] and Hyvärinen et al. [12], Sitnikova et al. [13], Chen [14] and Kato et al. [15] demonstrated in detail that the continuous Fourier transform and the discrete Fourier transform are not widely used in the process extraction of attributes due to their low efficiency. The most popular is the decomposition of the signal into harmonic components using the fast Fourier transform.

**2.2 EEG wavelet transform**

Wavelet transform is an integral transform, which is a convolution of a wavelet function with a signal. The wavelet transform transforms the signal from a temporal representation into a time-frequency representation. Khan et al. [16] and Balasubramanian et al. [17] and Khalaf et al. [18] and Kaleem [19] showed that the Fast Fourier Transform can be used with high precision in conjunction with the wavelet transform. The wavelet transform carries a huge amount of information about the signal, but, on the other hand, it has a strong redundancy, since each point in the phase plane affects its result.

**2.3 Principal component method**

Principal component analysis is a multivariate statistical analysis method used to reduce the dimension of the feature space with minimal loss of useful information. Electrooculography (EOG) is one of the most popular artifacts arising from pupil movement. In this case, the maximum amplitude of artifacts is observed in the frontal leads and decreases towards the occipital leads. An important point is that EOG artifacts are not associated with the current rhythm on the EEG and therefore must be removed from the signal. To remove artifacts caused by involuntary eye movements of the subject, a wide analysis of the main components is used with a multichannel EEG. Siuly et al. [20] and Artoni et al. [21] and Dea et al. [22] and Polat et al. [23] and Putilov et al. [24] have detailed this process. The method of principal components consists of finding for the initial data such as an orthogonal transformation into a new coordinate system. As a result, the directions of the basis vectors will be chosen in such a way that the covariance coefficient between the projections of the original dataset on different coordinate axes will ultimately be equal to zero, which eliminates the effect of pupil movement on the result.

**2.4 Other methods for feature extraction**

Some studies suggest non-standard solutions for extracting features from a signal. Singh et al. (2016) [25] proposed a new feature detection model based on the Hilbert-Huang transform, multivariate empirical mode decomposition (MEMD), and Hilbert transforms (HT). Comparisons are automatically made with the expansion of only the Fourier series and only on one model of the classification neural network. Cic et al. [26] has created three combined characteristics, combining spectral characteristics and time-frequency characteristics. The first combined function was a combination of all three types of functions used: RSD, IMFmed and IMFch. The second combination was RSD and IMFmed, and the third combination was RSD and IMFch. The efficiency criterion was a comparison with the Fourier series. Alomari et al. [27] proposed an automated computer platform for the classification of electroencephalographic (EEG) signals associated with the left and right-hand movements. The data was preprocessed using EEGLAB MATLAB tools, and artifact removal was done using AAR. Some works use several methods in an attempt to randomly combine. Jain et al.   [28] and Übeyli et al. [29] presented results that are only slightly superior to those obtained using Fourier spectral analysis.

## 3. Neural networks for EEG recognition tasks
### 3.1. Neural networks for diagnosing of diseases by EEG

Various brain diseases make changes in the normal picture of EEG, by which it is already possible to identify the nature of the disease, for example, to identify the location of the tumor or hemorrhage. The most popular studies in the field of neurological diagnosis are the work on the diagnosis of epileptic seizures. And it was here that thanks to machine learning, in the last decade, researchers managed to advance in the field of timely prediction of the occurrence of epilepsy. For example, the typical presence of epilepsy appears clearly on the EEG diagram (figure 3) presented by Cho et al.   [30], but it is much more difficult to predict it, that in fact more important.

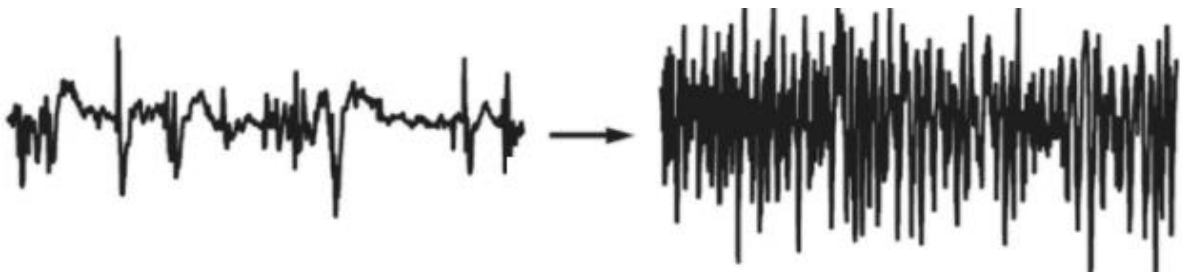

Fig.3. The moment of an epilepsy attack on the EEG diagram

Saba et al. [31] obtained encouraging results for solving the formulated problem — namely, the automatic detection of seizures and non-convulsive seizures by analyzing EEG signals. The model from the research consists of two parts: a convolutional neural network (CNN) module for recognizing local patterns and a module of a recurrent neural network (RNN).

Boonyakitanont et al. [32] proposed a deep CNN model for classifying seizures in multi-channel EEG signals. Model tests showed quite good results. But it is worth noting that the work had a standard approach - using the dataset, preprocessing and searching for the optimal neural network. Raj et al. [33], Yao et al. [34], Lopes et al. [35], Misiūnas et al. [36], and Khouma et al. [37] made similar researches for deciding this problem. They used a small dataset that was used and search from various machine learning models. The results of these studies are similar.

Research in the use of machine learning for the tasks of timely prediction of epilepsy attacks is very promising because about 55 million people suffer from epilepsy. At the same time, it is worth noting that today even the use of deep learning on microcontrollers (microcontrollers of the STM32 family, software CUBE.AI) is becoming available. From a technical point of view, the implementation of a portable device that will be able to report promptly about approaching based on data from a neural network does not present any difficulties.

Rodrigues et al. [38] used neural networks for recognizing diseases such as alcoholism, Bi et al. [39] for recognizing early Alzheimer's disease, Shim et al. [40] for detect schizophrenia disease, Bagheri et al. [41] for recognizing the inter-period epileptiform disease. Čukić et al. [42] used machine learning to detect depression. It has been experimentally proven that using the Higuchi fractal dimension, and selective entropy will reveal participants diagnosed with depression. Experimentally proved that using the Higuchi fractal dimension and selective entropy will reveal participants diagnosed with depression. Using neural networks for disease recognition saves time and, if there is a sufficient amount of dataset, it can become an alternative for a doctor, or as a tool for preliminary diagnostics. The problem lies in the fact that the trained model is not universal - each disease has separate weigh for the neural network, while a lot depends on the equipment, as a rule, images are used in recognition tasks - neural networks are very sensitive to expansion, so the picture must correspond to those images on which it was trained neural network. We can conclude that the field of application of neural networks for the recognition of diseases by EEG signals will only expand over time.

**3.2. Neural networks for lie recognition problems by EEG**

A few studies set the task of using EEG to recognize lies. Because for a subject it is extremely difficult to control the potential in the cerebral cortex, unlike signals from various physiological parameters sensors used in modern polygraphs. Therefore, a neural network is more likely to recognize a lie.

Yohan [43], used machine learning methods for analyzing EEG signals for recognizing lies is considered. To obtain the signals, a standard electrode placement system was used, but no information was provided on the type of electrode and related technical characteristics. To convert the EEG signals, the standard Fast Fourier Transform (FFT) method was used. In the research, that alpha waves carry information, the extraction of which will help to conclude whether the subject is telling the truth was shown. The author proposed comparing the power, the rms value (rms value) and the dispersion of the EEG signal. EEG graphs or references to this date are not presented in the work; it is not clear how the useful signal was extracted and interpreted.

Davatzikos et al. [44] received a high result to detect a lie. The accuracy estimated by cross-checking among participants not included in the training was 88%. At the same time, the parts of the brain were revealed, which are active in truth and the brain parts that are not active in a lie, figure 4-a. The author presents detailed graphs of the meaning of the "decisive function", which show whether the spatial activation pattern is representative of truth or falsehood, figure 4-b.

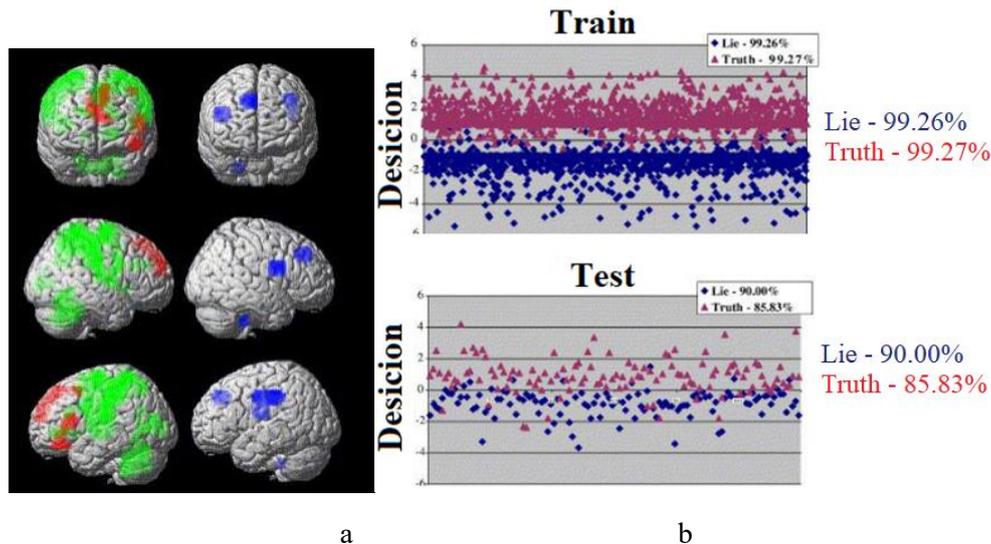

a            b

Fig. 4. Three-dimensional images of the brain, area with higher activity during the telling of the truth - green color, during the story of lying - red color. b - Graphs of the meaning of the "decisive function", which shows whether the subject is telling the truth or lying

Martinez-del-Rincon et al. [45], for same task used different models of neural networks, but result was similar.

The use of machine learning in the tasks of recognizing lies from a scientific point of view is interesting in obtaining the result about which parts of the brain were used and how the extraction of signs occurred. In the reviewed papers, as a rule, the results are presented rather briefly. It is not possible to judge the accuracy of the results. Since the minimum is necessary to make comparisons with the readings from a standard lie detector - polygraph.

This topic is quite local, but judging by the number of publications, it is of great interest. But even traditional devices for the analysis of lies are subject to some criticism, respectively, the use of a neural network for signal processing requires a large number of subjects, which was not observed in the previously described works, where the authors used several subjects.

### 3.3. Neural networks for research the physiology process by EEG

The tasks of physiology researchers include establishing a relationship between the physical activity of a person and potentials on the cortex of the brain. Alomari et al. [46] described the effect of fist rotation, leg movements and blinking for a short period of time on the EEG signal. The work uses the PhysioNetEEG dataset (https://physionet.org/). The waveforms obtained from the PhysioNet EEG dataset were analysed using wavelet analysis and machine learning. It is not clear how to use the trained model when using other types of electrodes, because in the work did not use the standard system for measuring electrodes 10-20. Batail et al [47], Chen et al [48], Panicker et al. [49] and Roy et al. [50] used similar methods and received the same conclusions.

Research in this direction is one of the most promising since the goal of these works is to manage various external mechanisms for people with limited possibilities. But today, it this field still are no tangible successes. A possible reason for this is that in most studies dry and wet electrodes were used. The method with surgical intervention and subcutaneous electrode placement not so popular, for obvious reasons.

In this case, great hopes are pinned on neural networks, since today, to record EEG signals, it is necessary for the subjects to be at rest, but the scope of application, for example, the brain-computer interface is much wider and includes such industries as exoskeleton control, robotic arm control, gaming industry where the active movement of the subject's limbs is necessary. In this case, there are a huge number of motor artifacts that must be properly cleaned up before using the signal for subsequent tasks.

### 3.4. Neural networks for control artifact in EEG signals

Pre-processing of image images of signals began to develop along with the first data obtained with EEG. Today, there is a very extensive and complex mathematical theory behind this, which includes various methods of pre-processing.

A few researchers suggest using neural networks as an instrument. From a technical point of view, this kind of implementation is extremely difficult. Since initially the cause of the artifacts is unknown and set the criteria for the neural network by which it will find out where the noise is and where the useful signal is not easy. Raj et al. [51] presented a brief overview of machine learning algorithms in the processing of EEG artefacts is given. Mayeli et al. [52] described the results of a comparison of a linear (linear discriminant analysis) and two nonlinear classifiers built based on neural networks. The result of the study is the conclusion that the use of non-linear classifiers does not far exceed the results of standard classification. Kang et al. [53], proved that muscle and residual visualization artefacts primarily affect the amplitude of the power spectrum in the beta frequency range. The purpose of eliminating artifacts is to obtain the following EEG diagram, figure 5.

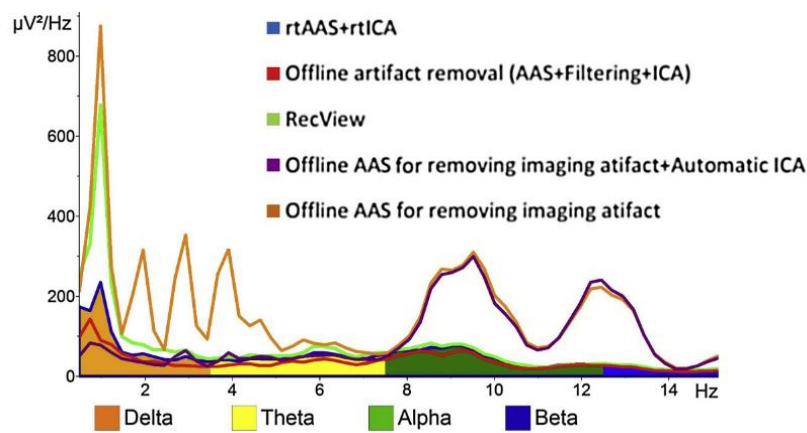

Fig.5. An example of the spectral power density of the O1 electrode (for subject 2) during runs with closed eyes using various algorithms for removing artefacts

In conclusion, we can add to the topic of artefacts in the EEG that the elimination of artefacts is not regulated in any way and the causes of artefacts are great and their nature and behaviour may depend on various external and internal factors. Islam et al [54], Levitt et al. [55], Yang et al. [56], Dhindsa et al. [57], and Abdulla et al. [58] presented the similar research with the different neural network model and its configuration. But the conclusions obtained as a result, as a rule, have a similar explanation.

**3.5. Neural networks for images classification tasks by EEG**

Good opportunities for studying EEG using a neural network are available during the training of neural networks when viewing various images by subjects while recording EEG signals. When the subject is viewing images it is difficult to predict which part of the brain and how will respond to the image, so we need to process the data from all the electrodes. Due to the large volume of the dataset, it is advisable to analyse them at the expense of neural networks.

Muller et al. [59] considered preliminary processing and classification methods for effective interaction of the brain and computer-based on EEG. To capture brain signals associated with motor tasks, EEG recorded 118 electrodes mounted on the scalp. Moreover, the study additionally recorded signals of electrooculogram (EOG) and electromyogram (EMG). The results of the study are quite difficult to interpret since the authors, as a result, did not make a clear distribution due to what indicators the neural network model made the forecast as a result.

Tsunoda et al. [60] proposed a new optical predictor, which uses a combination of a common spatial pattern network (CSP) and a short-term memory network (LSTM) to obtain an improved classification of EEG signals. Improvements in average erroneous classification are 3.09% and 2.07% for the BCI Competition IV I and GigaDB dataset, respectively.

Kang et al. [61] described in enough detail the problems of signal pre-processing. The paper considers the analysis of data sets of electroencephalography (EEG) and local field potential (LFP). It was found that cross-spectral density is a widely used metric for understanding the synchronous nature of a signal in different frequency bands. The cross-spectral density in the work was built by converting the time series into a frequency representation and then calculating the complex covariance matrix in each frequency band. Continued this study, Jiao et al. [62] presented an attempt to come to an understanding of how to read the image that a person sees from the cerebral cortex. As a result, the authors only presented recommendations in which direction it is advisable to move to solve this problem.

Studying the effect of viewing various images by a man and changing his EEG signals due to the receipt of a huge number of the dataset is ideal for using deep neural networks. At the same time, it is worth noting that it is still difficult to conclude the results obtained in the studies since initially, the researcher does not know what signals can be generated when viewing the picture. Therefore, the likelihood of accepting an artefact as a useful signal is high. The solution to this problem is to conduct many studies with different types of electrodes and in split rooms.

So far, from the point of view of practical results, there has been no progress in this case. As for visual perception and EEG response, a large number of EEG works on P300 [63, 64], SSVEP [65, 66] published where the authors managed to control robots and other objects, but on the other hand, it is much easier to do this using i-tracking which are not subject to motor artifacts.

### 3.6. Neural networks for sleep quality control by EEG

Rostamabad et al. [67] created an algorithm for analyzing the state of sleep on signals from the EEG. The algorithm is constructed as follows. Initially, the network operates by signal segmentation, and then the following network is used to classify a calm sleep and a nervous state during sleep into two classes. The author presents experimental studies that have shown the advantages of this approach over standard polysomnography (PSG). Only 2 electrodes were used in the work, too little data, and a small number of subjects. Chambon et al. [68], described in detail the sleep process and pre-processing of signals with EEG.

The study of the sleep process is interesting from the point of view of obtaining data in which the brain is less active as a comparison with the data when the brain is active, can help to identify several not typical of artifacts. This field is interesting for scientists studying human sleep and for monitoring meditation. [69, 70]

### 3.7. Theses in neural networks for EEG control

Several theses should be singled out separately because in them the difference from the manuscripts is quite actively described research processes. Shim et al. [71] provided a detailed description of the neural network development technique - ConvNet, which consists of four convolution-max-pool blocks, followed by three standard convolution-max-pool blocks and a dense softmax classification layer. Aljazaery et al. [72] used non-deep neural networks. The results obtained in the work do not exceed the accuracy described previously. Peker et al. [73] developed a recurrent neural network algorithm for decoding brain EEG. The paper presents signal processing for four moments - moving the hand to the left, moving the hand to the right, both hands moving and rest. The dataset was used to train it offline on the CPU and GPU using Theano packages. For training the model and its subsequent verification, a small amount of data was used, as a result, it is doubly difficult to evaluate the accuracy of the developed neural network. Chambayil et al. [74] and Song et al. [75] presented a description of a study on the topic of determining the processes occurring in an EEG using neural networks is. The work is interesting from a technical point of view.

### 3.8. Other researches involving machine learning for EEG analysis

Yuan et al. [76], proposed to use the ConvNets deep network for end-to-end EEG analysis. The study showed how to design and train ConvNets for decoding task-related information from unprocessed EEG without manually created functions. The result of the research revealed the potential of deep ConvNets in combination with visualization methods for EEG-based brain mapping.

David eet al. [77], expanded the classical nonlinear model of alpha rhythms with lumped parameters, originally developed for generation. Tests showed a higher accuracy of image classification with the proposed model. Akkar et al. [78], researched a neural network (DNN) in the context of decoding electroencephalographic signals for cognitively categorical information. Practical conclusions were drawn in the work: the use of too many neurons in the hidden layer can lead to the problem of excessive matching and other conclusions. Gandhi et al. [79] proposed a new architecture for processing neural information, inspired by quantum mechanics and incorporating the well-known Schrödinger wave equation. The proposed architecture, called recurrent quantum neural network (RQNN). The performance of RQNN was compared with both unfiltered EEG and well-known SGtechnique. The SG technique has been used as an approach to noise removal (thus, it is like RQNN) in biological signals such as ECG and EEG. The result obtained in the work is superior to the CNN scheme. The paper describes in quite some detail the preprocessing of the date and the neural network model itself. Song et al. [80] presented an attempt to use signals with EEG without preliminary processing. The authors presented a neural network model with convolutional layers with maximum integration. This model allowed to significantly reducing the percentage of objects with an accuracy of less than 70% (as a threshold for BCI). Kasabov et al. [81], proposed methodology is based on the recently proposed new diving neural network-NeuCub, the principle of which is well presented in figure 6.

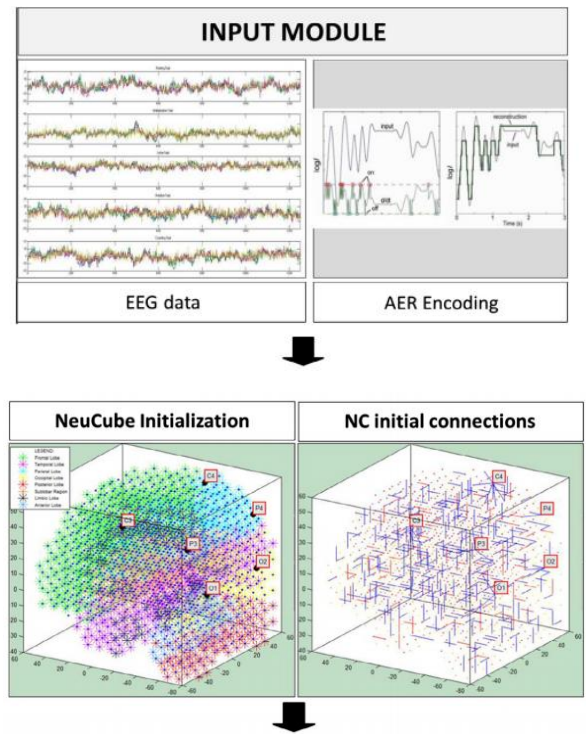

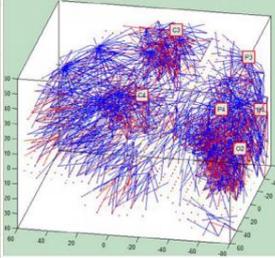

Fig.6. A graphical representation of the various stages of the proposed methodology- network- NeuCub

As you know, deep neural networks have shown better results than machine learning. The principle of operation of super-precise neural networks on the basis of which deep learning is built is that the input image is processed by weights and transformation due to convolution and pooling, and ultimately the data is sent to the activation functions. But in this work, a rather interesting approach is given where, after the first step of signal processing, the authors passed the obtained data through a cube in a three-dimensional format, which in the final count made the interpretation of the model even more difficult, but according to the authors, the result became better. This work shows that in the subject of EEG, one should not be content with the well-known models of neural networks, but new model can be created. DelPreto et al. [82] presented an interesting complex solution for providing the natural interaction of a person and a robot. Despite the presented conclusions in the study, it is very difficult to conclude how the results were obtained, since the study used the dataset from EEG and EMG.

## 4. Conclusion and discussion

We analyzed the processing of EEG signals for various industries using neural networks. In general, the main shortcomings were identified that do not allow fully feeling the quality of these works and making a correct comparison. The main problem of studying the use of the neural networks in the tasks of recognizing EEG signals is the lack of any standards to assess the quality of the presented model. As a result, each author in his own way evaluates the results obtained in the study. For the correct analysis of the results obtained during the study using machine learning and EEG data, the primary task is the correct use of electrodes, which allows getting rid of noise and other artifacts. We recommend mentioning in the manuscripts the next information about the conduct of the following procedures. Ignoring this can lead to inaccuracies in the operation of the neural network.

Recommended procedures for EEG researchers:

- Today wet electrodes are a recognized standard, in works in which the development of a dry electrode appears, it is necessary to make a comparison with wet electrodes. In this case, it is necessary to use the standard (ANSI / AAMI) EC-12: 2000 standard for disposable electrodes;

- It is necessary to use laboratory analyzers of impedance and amplitude-phase characteristics. For example, Solartron 1260 with a nominal accuracy of 0.1% from 5 ohms to 100 kOhms;

- It is necessary to use an electrochemical interface, for example, solartron 1287A. In which it is advisable to use the following functions of the device:

- galvanostat is a device that maintains a constant current in a specific cell, regardless of the difference in electrode potential;

- potentiostat is an electronic device whose purpose is to automatically control the potential of the electrode and support a predetermined value of this electrode;

- It is necessary to measure the internal noise of the electrode;

- Need to refer to the standard of medical equipment (EN 60601 -- 1 -- 2), which describes tests for resistance to the power line frequency (50 Hz), magnetic field test (EN 61000 -- 4 -- 8) and resistance to Radio Frequency Test (RF) emissions (EN 61000-4 -- 3);

- Before research it is necessary to verify that the electrode has: immunity to frequency (50 Hz) in the power line, immunity to radiation of radio frequencies, immunity to radiated radio frequencies.

- Before testing, it is necessary to provide information about the patient's skin, resistance, and impedance;

- The presence of artifacts, the analysis of the obtained EEG is expediently carried out in conjunction with a neurophysiologist.

Despite the huge interest in the topic of using neural networks in the process of signal processing, it is still not possible to establish a direct relationship between external extraneous physical effects on the body and potential on the cerebral cortex. The results obtained in the studies, for the most part, can be useful as processing EEG signals, searching for various diseases and processing artifacts, but the stated goal of most studies, namely the online processing of EEG signals to control external objects, has not been achieved with dry or wet electrodes.

In the papers described earlier, no links were provided to the Kaggle site, which is the best place for workability analysis developed by the wide network models. For example, on the Kaggle, today there is a dataset of 39 datasets in the EEG theme. Competitions will help to give a real picture of the evaluation of machine learning models.

Perspective direction for the next research is the use of the Internet of things technology. STM32 microcontrollers with Esp8266 can work as in the Internet of things format and transfer their data directly to the cloud, for example through the ThingSpeak application and the Internet of Things API for storing and retrieving data from things using HTTP and MQTT protocols.

A promising direction for the development of the developed installation is the use of X-CUBE-AI software - AI. This extension can work with various deep learning environments such as Caffe, Keras, TensorFlow, Caffe, ConvNetJs, etc. Thanks to this, the neural network can be trained on a desktop computer with the ability to calculate on a GPU. After integration, use the optimized library for the 32-bit microcontroller STM32, which will allow online monitoring of subject state by EEG sensors and prevent the occurrence of epileptic shocks. Moreover, for this application, it is enough to consider only the Alpha range and enough 4 sensors of EEG sensors

**Conflicts of Interest**: None

**Funding**: None

**Ethical Approval**: Not required

**References**

1. Neelam, R. (2014). Analysis and Classification Technique Based On ANN for EEG Signals. Neelam Rout, International Journal of Computer Science and Information Technologies, 5 (4), 5103-5105

2. Ravan, M. (2019). A machine learning approach using EEG signals to measure sleep quality. AIMS Electronics and Electrical Engineering, 3, 347–358

3. Amorim, E., Stoel, M.,& Nagaraj, N. (2019). Quantitative EEG reactivity and machine learning for prognostication in hypoxic-ischemic brain injury, Clinical Neurophysiology, 130, Issue 10, 1908-1916

4. Congedo, M., Barachant, A., & Bhatia, R. (2017). Riemannian geometry for EEG-based brain-computer interfaces; a primer and a review. Brain-Computer Interfaces, 4, Issue 3, 155-174

5. Sharmila, A. (2018). Epilepsy detection from EEG signals: a review. Journal of Medical Engineering & Technology, 42, Issue 5, 368-380

6. Usman, S., & Khalid, S. (2019). Using scalp EEG and intracranial EEG signals for predicting epileptic seizures: Review of available methodologies. Seizure, 71, 258-269

7. Clarke, S., & Karoly, P. (2019). Computer-assisted EEG diagnostic review for idiopathic generalized epilepsy. Epilepsy & Behavior, DOI: 10.1101/682112

8. Al-Fahoum. (2014). Methods of EEG Signal Features Extraction Using LinearAnalysis in Frequency and Time-Frequency Domains. SRN Neuroscience, 7, 14-19

9. Mane, A., Biradar, S., & Shastri, K. (2019). Review paper on Feature Extraction Methods for EEG Signal Analysis. International Journal of Emerging Trend in Engineering and Basic Sciences, 2349-6967

10. Raut, A. (2010). EEG feature selection using mutual information and support vector machine: A comparative analysis. 32nd Annual International Conference of the IEEE EMBS Buenos Aires, Argentina, 4946-4947

11. Al-Salman, W., Li,Y. & Wen, P. (2019). Detecting sleep spindles in EEGs using wavelet fourier analysis and statistical features. Biomedical Signal Processing and Control, 48, 80-92

12. Hyvärinen, A., Ramkumar, P., Parkkonen, L., & Hari, R. (2010). Independent component analysis of short-time Fourier transforms for spontaneous EEG/MEG analysis. NeuroImage, 49, Issue 1, 257-271

13. Sitnikova, E., Hramov, A., Koronovsky, A., & Luijtelaar, G. (2009). Sleep spindles and spike–wave discharges in EEG: Their generic features, similarities and distinctions disclosed with Fourier transform and continuous wavelet analysis. Journal of Neuroscience, 180, Issue 2, 304-

14. Chen, G. (2014). Automatic EEG seizure detection using dual-tree complex wavelet-Fourier features. Expert Systems with Applications, 41, Issue 5, 2391-2394

15. Kato, K., Takahashi, K., Nobuaki, M., & Ushiba, J. (2018). Online detection of amplitude modulation of motor-related EEG desynchronization using a lock-in amplifier: Comparison with a fast Fourier transform, a continuous wavelet transform, and an autoregressive algorithm. Journal of Neuroscience Methods, 293, 289-298

16. Khan, H., Shanir, P., & Farooq, O. (2020). A hybrid Local Binary Pattern and wavelets based approach for EEG classification for diagnosing epilepsy. Expert Systems with Applications, 140, 115-120

17. Balasubramanian, G., Kanagasabai, A., & Mohan, J. (2018). Music induced emotion using wavelet packet decomposition—An EEG study. Biomedical Signal Processing and Control, 42, 115-128

18. Khalaf, A., Sejdic, E., & Akcakaya, M. (2019). Common spatial pattern and wavelet decomposition for motor imagery EEG- fTCD brain-computer interface. Journal of Neuroscience Methods, 320, 15, 98-106

19. Kaleem, M., Guergachi, A., & Krishnan, S. (2018). Patient-specific seizure detection in long-term EEG using wavelet decomposition. Biomedical Signal Processing and Control, 46, 157-165

20. Siuly, S., & Li, Y. (2015). Designing a robust feature extraction method based on optimum allocation and principal component analyze. Computer Methods and Programs in Biomedicine, 119, Issue 1, 29-42

21. Artoni, F., Delorme, A., & Makeig, S. (2018). Applying dimension reduction to EEG data by Principal Component Analysis reduces the quality of its subsequent Independent Component decomposition. NeuroImage, 175, 176-187

22. Dea, F., & Stecca, M. (2019). A Big-Data-Analytics Framework for Supporting Classification of ADHD and Healthy Children via Principal Component Analysis of EEG Sleep Spindles Power Spectra. Procedia Computer Science, 159, 1584-1590

23. Polat, K., & Güneş, S. (2008). Artificial immune recognition system with fuzzy resource allocation mechanism classifier, principal component analysis and FFT method based new hybrid automated identification system for classification of EEG signals. Expert Systems with Applications, 34, Issue 3, 2039-2048

24. Putilov, A. (2015). Principal component analysis of the EEG spectrum can provide yes-or-no criteria for demarcation of boundaries between NREM sleep stages. Sleep Science, 8, Issue 1, 16-23

25. ingh, M., & Goyat, R. (2016). Feature Extraction for the Analysis of Multi-Channel EEG Signals Using HilbertHuang Technique. International Journal of Engineering and Technology, 8, 17-27

26. Cic, M. (2019). Optimal set of EEG features in infant sleep stage classification. Turkish Journal of Electrical Engineering & Computer Sciences, 27, 605-615

27. Alomari, M., Samaha, A., & Kamha, K. (2013). Automated Classification of L/R Hand Movement

EEG Signals using Advanced Feature Extraction and Machine Learning. (IJACSA) International Journal of Advanced Computer Science and Applications, 4, 207-212

28. Jain, S., Dye, T., & Kedia, P. (2019). Value of combined video EEG and polysomnography in clinical management of children with epilepsy and daytime or nocturnal spells. Seizure, 65, 1-5

29. Übeyli, E. (2008). Analysis of EEG signals by combining eigenvector methods and multiclass support vector machines. Computers in Biology and Medicine, 38, Issue 1, 14-22

30. Cho, K., & Jang, H. (2020). Comparison of different input modalities and network structures for deep learning-based seizure detection, Scientific Reports, 10, 122

31. Saba, M. (2013). A Review of EEG-Based Brain-Computer Interfaces as Access Pathways for Individuals with Severe Disabilities. Assistive Technology. The Official Journal of RESNA, 25, 99-110

32. Boonyakitanont, P., Chomtho, K., & Songsiri, J. (2019). A Comparison of Deep Neural Networks for Seizure Detection in EEG Si, https://doi.org/10.1101/702654

33. Raj, K., Rajagopalan, S., & Bhardwaj, S. (2018). Machine learning detects EEG microstate alterations in patients living with temporal lobe epilepsy. Seizure, 61, 8-13

34. Yao, L., Cai, M., & Guo, Y. (2019). Prediction of antiepileptic drug treatment outcomes of patients with newly diagnosed epilepsy by machine learning. Epilepsy & Behavior, 96, 92-97

35. Lopes, M., Junges, L., & Tait, L. (2019). Computational modelling in source space from scalp EEG to inform presurgical evaluation of epilepsy. Clinical Neurophysiology, 131(1), 225-234

36. Misiūnas, A., Meškauskas, T., & Samaitienė, R. (2019). Algorithm for automatic EEG classification according to the epilepsy type: Benign focal childhood epilepsy and structural focal epilepsy. Biomedical Signal Processing and Control, 48, 118-127

37. Khouma, O., & Diop, I. (2019). Novel Classification Method of Spikes Morphology in EEG Signal Using Machine Learning. Procedia Computer Science, 148, 70-79

38. Rodrigues, J., Filho, P., & Peixoto, E. (2019). Classification of EEG signals to detect alcoholism using machine learning techniques. Pattern Recognition Letters, 125, 140-149

39. Bi, X., & Wang, H. (2019). Early Alzheimer's disease diagnosis based on EEG spectral images using deep learning. Neural Networks, 114, 119-135

40. Shim, M., Hwang, H., & Lee, S. (2016). Machine-learning-based diagnosis of schizophrenia using combined sensor-level and source-level EEG features. Schizophrenia Research, 176, Issues 2–3, 314-319

41. Bagheri, E., & Jin, J. (2019). A fast machine learning approach to facilitate the detection of interictal epileptiform discharges in the scalp electroencephalogram. Journal of Neuroscience Methods, 326, 156-160

42. Čukić, M., Pokrajac, D., & Stokić, M. (2018). EEG machine learning with Higuchi's fractal dimension and Sample Entropy as features for successful detection of depression, DOI 10.1007/s11571-020-09581-x

43. Yohan, K. (2019). Using EEG and Machine Learning to perform Lie Detection, DOI 10.1007/s11571-020-09581-x

44. Davatzikos, C., Ruparel, K. & Fan,Y. (2005). Classifying spatial patterns of brain activity with machine learning methods: Application to lie detection. NeuroImage, 28, Issue 3, 663-668

45. Martinez-del-Rincon, J., Martinez-del-Rincon, M., & Xavier, T. (2017). Non-linear classifiers applied to EEG analysis for epilepsy seizure detection. Expert Systems with Applications, 86, 99-112

46. Alomari M., AbuBaker, A., Aiman Turani, A., Baniyounes, M., & Manasreh, A. (2014). EEG Mouse:A Machine Learning-Based Brain Computer Interface. (IJACSA) International Journal of Advanced Computer Science and Applications, 5, 193-198

47. Batail, J., Bioulac, S., Cabestaing, F., & Daudet, C. (2019). EEG neurofeedback research: A fertile ground for psychiatry? L'Encéphale, 45, Issue 3, 245-255

48. Chen, L., Zhao, Y., & Zhang, J. (2015). Automatic detection of alertness/drowsiness from physiological signals using wavelet-based nonlinear features and machine learning. Expert Systems with Applications, 42, Issue 21, 7344-7355

49. Panicker, S., & Gayathri, P. (2019). A survey of machine learning techniques in physiology based mental stress detection systems. Biocybernetics and Biomedical Engineering, 39, Issue 2, 444-469

50. Roy, R., Charbonnier, S., & Bonnet, S. (2014). Eye blink characterization from frontal EEG electrodes using source separation and pattern recognition algorithms. Biomedical Signal Processing and Control, 14, 256-264

51. Raj, K., Rajagopalan, S., Bhardwaj, S., & Panda, R. (2018). Machine learning detects EEG microstate alterations in patients living with temporal lobe epilepsy, Seizure, 61, October 2018, 8-13

52. Rodrigues, J., Filho, P., & Peixoto, E. (2019). Classification of EEG signals to detect alcoholism using machine learning techniques. Pattern Recognition Letters, 125, 140-149

53. Mayeli, A., Zotev, V., & Refai, H. (2016). Real-time EEG artifact correction during fMRI using ICA. Journal of Neuroscience Methods, 274, 27-37

54. Islam, M., M.El-Hajj. A., & Alawieh, H. (2020). EEG mobility artifact removal for ambulatory epileptic seizure prediction applications. Biomedical Signal Processing and Control, 55, 120-125

55. Levitt, J., Nitenson, A., Koyama, S., & Heijmans, L. (2018). Automated detection of electroencephalography artifacts in human, rodent and canine subjects using machine learning. Journal of Neuroscience Methods, 307, 53-59

56. Yang, B., Duan, K., & Fan, C. (2018). Automatic ocular artifacts removal in EEG using deep learning. Biomedical Signal Processing and Control, 43, 148-158

57. Dhindsa, K. (2017). Filter-Bank Artifact Rejection: High performance real-time single-channel artifact detection for EEG. Biomedical Signal Processing and Control, 38, 224-235

58. Abdulla, S., Diykh, M., & Laft, R. (2019). Sleep EEG signal analysis based on correlation graph similarity coupled with an ensemble extreme machine learning algorithm. Expert Systems with Applications, 138, 230-237

59. Muller, K., Tangermann, M., & Dornhege, G. (2008). Machine Learning for Real-Time Single-Trial EEG Analysis: From Brain-Computer Interfacing to Mental State Monitoring, Journal of Neuroscience Methods, 167(1), 82-90

60. Tsunoda, T., Sharma, A., & Tsunoda, T. (2019). Brain wave classification using long short-term memory network based OPTICAL predictor. Scientific Reports, 9, 140-145

61. Kang, G., Jin, S., & Kang, W. (2018). EEG artifacts removal using machine learning algorithms and independent component analysis. Clinical Neurophysiology, 129, 24-27

62. Jiao, Z., & You, H. (2008). Decoding EEG by Visual-guided Deep Neural Networks. Proceedings of the Twenty-Eighth International Joint Conference on Artificial Intelligence, 1387-1393

63. Li, M., Yang, G., & Li, H. Effect of the concreteness of robot motion visual stimulus on an event-related potential-based brain-computer interface. Neuroscience Letters Volume 720, 16 February 2020, 134752

64. Mouli, S., & Palaniappan, R. (2020). DIY hybrid SSVEP-P300 LED stimuli for BCI platform using EMOTIV EEG headset. HardwareX, 8, e00113

65. Chai, X., Zhang, X., Guan, K., Lu, Y., Liua, G., & Zhang, T. (2020). A hybrid BCI-controlled smart home system combining SSVEP and EMG for individuals with paralysis. Biomedical Signal Processing and Control, 56, 101687

66. Erkan, E., & Akbaba, M. (2018). A study on performance increasing in SSVEP based BCI application. Engineering Science and Technology, an International Journal, 21, Issue 3, 421-427

67. Rostamabad, A., Reilly, J., & Hasey, G. (2013). A machine learning approach using EEG data to predict response to SSRI treatment for major depressive disorder, Clinical Neurophysiologym 124, Issue 10, 1975-1985

68. Chambon, S., Thorey, V., & Arnal, P. (2019). A deep learning approach to detect multiple sleep micro-events in EEG signal. Journal of Neuroscience Methods, 321, 64-78

69. Colgan, D., Memmott, T., Klee, D., Ernst, L., & Han, S. (2019). A single case design to examine short-term intracranial EEG patterns during focused meditation. Neuroscience Letters, 711, 15, 134441

70. Stapleton, P., Dispenza, J., & McGill, S. (2020). Large effects of brief meditation intervention on EEG spectra in meditation novices. IBRO Reports, 9, 290-301

71. Shim, M., Hwang, H., & Kim, D. (2016). Machine-learning-based diagnosis of schizophrenia using combined sensor-level and source-level EEG features, Schizophrenia Research, 176, Issues 2–3, 314-319

72. Aljazaery, I., & Abduladhem, A. (2017). Classification of Electroencephalograph (EEG) Signals Using Quantum Neural Network. Signal Processing, An International Journal (SPIJ), 4, Issue 6, 123-134

73. Peker, M. (2016). An efficient sleep scoring system based on EEG signal using complex-valued machine learning algorithms, Neurocomputing, 207, 165-177

74. Chambayil, B., & Jha, R. (2010). EEG Eye Blink Classification Using Neural Network. Proceedings of the World Congress on Engineering, ISBN: 978-988-17012-9-9

75. Song, Y., Crowcroft, J., & Zhang, J. (2012). Automatic epileptic seizure detection in EEGs based on optimized sample entropy and extreme learning machine. Journal of Neuroscience Methods, 210, Issue 2, 132-146

76. Yuan, Q., Zhou, W., & Li, S. (2011). Epileptic EEG classification based on extreme learning machine and nonlinear features. Epilepsy Research, 96, Issues 1–2, 29-38

77. David, O., & Friston, K. (2003). A neural mass model for MEG/EEG: coupling and neuronal dynamics. NeuroImage, 20, 1743–1755

78. Akkar, H., & Jasim, F. (2018). Intelligent Training Algorithm for Artificial Neural Network EEG Classifications. International Journal of Intelligent Systems and Applications, 10(5), 33-41

79. Gandhi, V., Coyle, D., & Prasad, G. (2014). Quantum Neural Network-Based EEG Filtering for a Brain-Computer Interface. EEE Transactions on Neural Networks and Learning Systems, 25(2), 278-288 ·

80. Song, J., Hu, W., & Zhang, R. (2016). Epileptic EEG classification based on extreme learning machine and nonlinear features. Neurocomputing, 175, 383-391

81. Kasabov, N., & Capecci, E. (2015). Spiking neural network methodology for modelling, classification and understanding of EEG spatio-temporal data measuring cognitive processes. Information Sciences, 294, 565–575

82. DelPreto, J., & Salazar-Gomez, A. (2018). Plug-and-Play Supervisory Control Using Muscle and Brain Signals for Real-Time Gesture and Error Detection, Conference: Robotics: Science and Systems 2018, At Pittsburgh, Pennsylvania, SBN 978-0-9923747-4-7